# Topologically constrained high intensity light propagation in air

A. Goffin[1,2], L. Railing[1,3], G. Babic[1,3], and H. M. Milchberg[1,3,4*]

[1]*Institute for Research in Electronics and Applied Physics, University of Maryland, College Park, MD 20742*
[2]*Los Alamos National Laboratory, Los Alamos, NM 87545*
[3]*Dept. of Physics, University of Maryland, College Park, MD 20742*
[4]*Dept. of Electrical and Computer Engineering, University of Maryland, College Park, MD 20742*

[*]*Corresponding author: milch@umd.edu*

**Abstract**

We experimentally demonstrate how spatiotemporal optical vortices (STOVs) control long-range atmospheric filamentation of intense laser pulses. High-power pulses long enough to overlap with the delayed rotational nonlinearity of air molecules ($> \sim 100$ fs) undergo periodic collapse arrest events, each of which generates toroidal STOV pairs with $\pm 1$ topological charge that separate and accumulate into increasingly squeezed arrays of $+1$ charges at the front of the pulse and $-1$ charges at the back. These dynamics manifest as periodic energy deposition peaks along the propagation path and a pulse envelope modulated into a temporal intensity comb. Filamentation in this regime can be understood in terms of self-organized, topologically constrained defect dynamics embedded within nonlinear wave propagation.

When a sufficiently high-peak-power femtosecond pulse undergoes nonlinear self-focusing in a uniform transparent medium, the process leads to runaway beam collapse: rapid shrinkage of the beam width and growth in peak intensity until arrested by plasma generation and defocusing [1–3]. The dynamic interplay between self-focusing and plasma defocusing leads to long-distance propagation of a narrow, high intensity "core" region surrounded by a wider "reservoir" which exchanges power with it [4]. Such propagating optical structures, called filaments, enable multiple applications [5–12] that depend on long-range delivery of high peak and average power.

While the "dynamic interplay" aspect of filamentation—the complex interaction of multiple competing transient nonlinear effects—has been studied with extensive simulations [1,2,4,13], we have recently developed a high-level, unifying description that offers deep physical insight while explaining and predicting complex behaviour [14]. In both the non-relativistic and relativistic regimes, filament evolution can be understood in terms of spatiotemporal optical vortex (STOV)-directed intrapulse electromagnetic energy flow. STOVs form due to the extreme spatiotemporal phase shear generated during self-focusing collapse arrest [14,15], and quickly evolve into toroidal vortex rings that wrap around the pulse axis, imposing topologically mandated energy flows that mediate nonlinear propagation. STOV dynamics explains well-known phenomena such as refocusing cycles [14], pulse splitting [4,16,17] and X-wave formation [18,19]. To date, however, no experiments have directly connected STOV dynamics with their signature effects on pulse envelope evolution and long-distance propagation. In addition to its fundamental physics interest, such an understanding is crucial to optimization of long range filament applications such as remote detection [5,6,10], the guiding of lightning [7], and air waveguides [8,9,20,21]

In this paper, we show with experiments and simulations that propagation and evolution of femtosecond filaments in air is governed by repeated generation of toroidal STOV pairs with $\pm 1$ topological charge. The $+1$ charges migrate forward and accumulate into an array of toroidal STOVs at the front of the pulse, while the $-1$ charges accumulate at the back. These two arrays direct intrapulse electromagnetic energy flow, drive a regular and predictable multi-peak pulse envelope structure, and cause periodic modulations of the energy deposition profile along the propagation track. For short pulses that sample mainly the instantaneous electronic response, the periodic axial deposition and envelope modulations go away and the filament propagation range is reduced. The delayed rotational response of air molecules regularizes filament propagation dynamics, making it less sensitive to fluctuations in pulse intensity. These dynamics are exploited to generate greatly extended filaments in air.

The filamentation process initiates when self-focusing overcomes diffractive beam spreading for laser peak power $P > P_{cr} = 3.77\lambda^2/8\pi n_{2,eff} n_0$ [22,23], where $n_0$ is the medium's unperturbed refractive index and $n_{2,eff}$ is its effective nonlinear index of refraction. In air, the rotational response of the constituent N$_2$ and O$_2$ molecules contributes significantly to $n_{2,eff}$ for sufficiently long pulses. Pulses shorter than ~60 fs mainly drive the near-instantaneous electronic nonlinearity from randomly oriented molecules; each time slice of the pulse primarily focuses itself. For longer pulses, the molecules increasingly align with the laser polarization until the nonlinearity is dominated by the rotational response [24,25]; focusing by the induced "molecular lens" is delayed by a characteristic rotational timescale of several hundred femtoseconds. In air, $n_{2,eff}$ is therefore pulsewidth dependent and $P_{cr}$ ranges over $\sim 2.5 - 12$ GW at $\lambda = 800$ nm [24].

In our experiments, we examine the pulsewidth-dependent laser field evolution and propagation track while keeping the ratio $P/P_{cr}$ constant. We must therefore find the correct $n_{2,eff}$ for a pulse of width $\tau$ that embodies the electronic plus rotational responses over the full pulse evolution. As shown in [26],

$$n_{2,eff}(\tau) = \left(\int_{-\infty}^{\infty} I^2(t)dt\right)^{-1} \int_{-\infty}^{\infty} I(t)\Delta n_b(t)dt, \qquad (1)$$

where $\Delta n_b(t) = n_2 I(t) + \int_{-\infty}^{t} dt' R(t-t') I(t')$ is the nonlinear refractive index shift contributed by bound electrons. Here, the first term is the near-instantaneous electronic response and the second term is the delayed molecular rotational response, where $R(t)$ is the diatomic molecular response function and $I(t)$ is the laser intensity envelope [24,25,27]. This can be written as $n_{2,eff}(\tau) = n_2 + n_{2,rot}(\tau)$, separating the prompt and delayed bound contributions. Figure 1(a) plots $n_{2,eff}(\tau)$ vs. $\tau$ for N$_2$, O$_2$ and air (80% and 20% N$_2$ and O$_2$ contributions); the result is dependent only on pulse duration and not peak intensity, with the peak value near $\tau \sim 300$ fs. Note that for $\tau \to 0$, $n_{2,rot}(\tau) \to 0$ and $n_{2,eff}(\tau) \to n_2$.

Our experiments used a Ti:Sapphire chirped pulse amplification (CPA) laser with $\lambda_0 = 812$ nm and minimum FWHM pulse duration $\tau = 45$ fs. The pulsewidth was adjusted over the range $45$ fs $- 2$ ps by scanning the CPA compressor. After compression, the pulse was down-collimated with an off-axis reflective telescope to a spot size $w_0 \sim 1.4$ mm, with a slight $\sim f/1500$ focus to stabilize the collapse location [28] and limit filament length to lab scales. The shot-to-shot energy of the pulse was measured with a calibrated CCD camera. The pulse energy was varied to maintain $P/P_{cr}(\tau) = 5.5$ as the pulsewidth was varied; the required power and energy are plotted in Fig. 1(b). For the longest pulse in the scan, $\tau = 2$ ps, $\varepsilon_{pulse} \sim 30$ mJ.

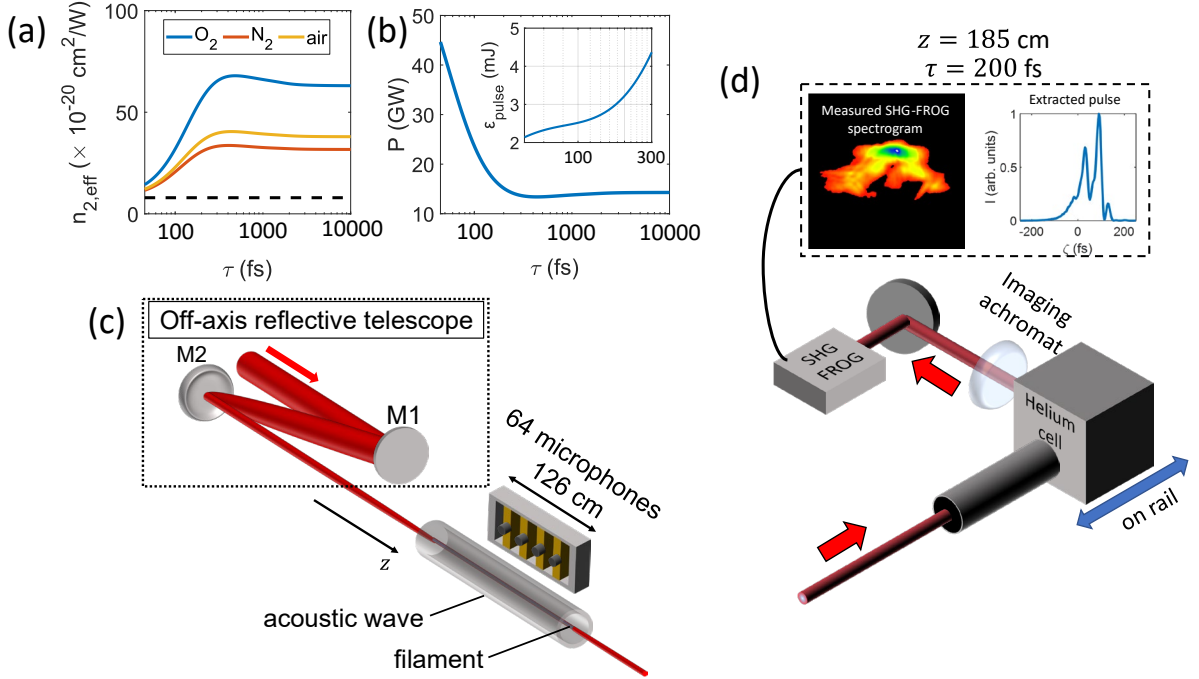

**Figure 1. (a)** $n_{2,eff}(\tau)$ vs. $\tau$ for a Gaussian pulse for 1 atm of $O_2$, $N_2$, and air starting from $\tau = 45$ fs. The black dashed line corresponds to the electronic nonlinear index of air, $n_2 \approx 7.9 \times 10^{-20}$ cm$^2$/W [24]. **(b)** Pulse peak power vs. $\tau$ required to maintain $P/P_{crit}(\tau) = 5.5$. *Inset:* Associated pulse energies from 45 fs to 300 fs. **(c)** Filament generation and single-shot microphone array configuration. Mirrors M1 ($f = +75$ cm) and M2 ($f = -25$ cm) form a 3× down-collimating telescope. The 126-cm-long microphone array consists of 64 microphones with 2-cm spacing, each with 24-bit A/D converter at 44.1 kHz, synchronized through a microcontroller hub. **(d)** Rail-mounted single-shot SHG-FROG measures envelope and phase of a filamenting pulse mid-flight at the air-helium interface.

The mid-flight pulse envelope and phase during propagation was measured directly by propagating the filament into the entrance of a slow outflow helium cell [15,29] (see Fig. 1(d)), where it terminates at the air-helium interface owing to $n_2^{helium} < 0.05\, n_{2,eff}$ [24,30]. The interface was then relay imaged, using an aperture to select the filament core, to a second harmonic generation frequency-resolved optical gating (SHG-FROG) diagnostic [31]. The helium cell and SHG-FROG were mounted to a carriage positioned along the filament propagation path, enabling mid-flight measurements of the pulse during propagation.

The full propagation track of the filament was measured by single-shot sonographic imaging of the filament-induced single-cycle acoustic wave using a synchronized microphone array placed 3 mm from the propagation axis [9,28]. Shot-to-shot variation of the initial self-focusing collapse location from air turbulence and pulse energy fluctuations makes single-shot measurement essential [28], and enables multi-shot averaging. The amplitude of the cylindrical acoustic wave launched by the filament at axial location $z$ is proportional to the local energy deposited per unit length, $\partial_z \varepsilon_{dep}$ [32], whose axial distribution is measured by the microphone array. Femtosecond filaments deposit energy in air mainly through optical field ionization plus rotational excitation of $O_2$ and $N_2$ molecules [33], with electron-neutral and electron-ion collisional heating negligible for the < 2 ps pulsewidths of our experiments. For full filament tracks ranging to > 2 m, microphone array signals were stitched together. The microphone array was calibrated as described in the supplementary material [26]. Physical insight is provided by propagation simulations with the code YAPPE [26], an implementation of the unidirectional pulse propagation equation [34]. YAPPE includes the electronic, rotational and plasma nonlinear responses of air constituents, and calculates filament energy deposition from field ionization, molecular rotational excitation, and collisions.

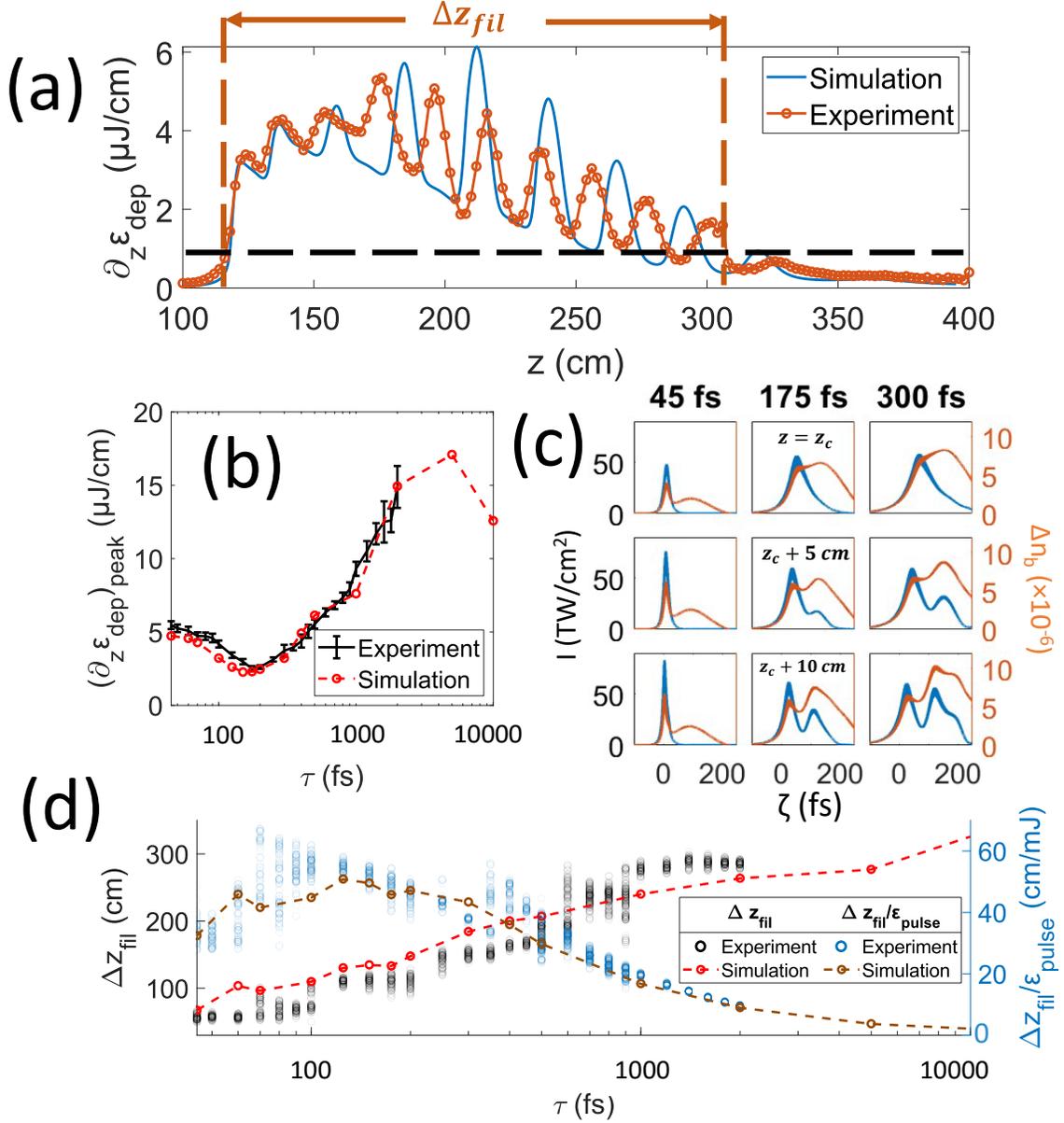

**Figure 2. (a)** Energy calibrated microphone trace (stitched 100 shot average, with single-shot traces first aligned to common collapse position $z_c$) for pulsewidth $\tau = 500$ fs, along with YAPPE simulation of energy deposition. The dashed black horizontal line marks the deposition threshold $\partial_z \varepsilon_{dep} = 0.7$ μJ/cm, which we use to define the filament length $\Delta z_{fil}$. **(b)** Measured $(\partial_z \varepsilon_{dep})_{peak}$ vs. pulsewidth $\tau$. Error bars are the standard deviation over 100 shots. Red circles mark $(\partial_z \varepsilon_{dep})_{peak}$ from YAPPE simulations. **(c)** Simulated on-axis intensities $I(r = 0, \zeta; z)$ and bound electron refractive index shift $\Delta n_b(\zeta; z)$ ($\zeta = t - z/v_g$) for pulse durations $\tau = 45, 175,$ and $300$ fs, for three locations after collapse arrest. **(d)** Black circles: filament length $\Delta z_{fil}$ vs. pulsewidth. Each circle corresponds to $\Delta z_{fil}$ for a single stitched trace, 100 traces per pulse duration. Blue circles: Filament length per unit energy for each pulse duration. All points are for $P/P_{cr} = 5.5$. Red and brown circles and lines: YAPPE simulation results.

Our results reveal an intimate connection between the axial energy deposition profile of the filament, its temporally modulated pulse envelope, and STOVs. One of the striking results of our experiments is the distinctive, periodically modulated energy deposition profiles observed for filamenting pulses longer than ~150 fs. Figure 2(a) plots the stitched and 100-shot-averaged microphone trace for a $\tau = 500$ fs pulse ($\varepsilon_{pulse} = 7.2$ mJ). Axial position $z$ is referenced to mirror M2 in Fig. 1(c). Starting with beam collapse at $z \sim 115$ cm, the experimental and simulated deposition peaks are similar in onset, spacing, and amplitude, with differences stemming from simulation sensitivity to model details [26]. The regular peak spacing (of ~$20 - 25$ cm) is caused by periodic refocusing and collapse arrest events induced by the molecular lens. The peak spacing is explained by a simple model for periodic refocusing [26].

We first consider a single refocusing event by examining the effect of varying pulsewidth on energy deposition. Figure 2(b) plots the peak energy deposition $(\partial_z \varepsilon_{dep})_{peak}$ along the filament track vs. $\tau$, with the results well-matched by simulations. These results are understood from simulations of the on-axis laser intensity $I(r = 0, \zeta; z)$ and transient refractive index $\Delta n_b(r = 0, \zeta; z)$ plotted in Fig. 2(c) for 3 different pulsewidths for several axial locations after collapse arrest at $z = z_c$. Here $\zeta = t - z/v_g$ is local time in a window moving at the group velocity $v_g$ of the initial pulse, and $z$ is the location of the moving window. For $\tau = 45$ fs, the near-instantaneous electronic response dominates, with the highest power time slices experiencing the strongest self-focusing and causing higher energy deposition. For $\tau = 175$ fs, the now larger rotational response lags the peak of the pulse; molecular lens focusing of the lower power time slices results in less energy deposited. For $\tau = 300$ fs, the rotational response grows further and overlaps more with high-power time slices, leading to stronger focusing and energy deposition. As $\tau$ increases further, an increasing number of refocusing events occurs, with $(\partial_z \varepsilon_{dep})_{peak}$ increasing through the maximum experimental pulsewidth $\tau = 2$ ps, with simulations predicting a subsequent rollover as increasing avalanche ionization limits collapse arrest to lower intensities, causing lower peak deposition. The dynamics leading to continually increasing peak deposition for $\tau > 300$ fs are controlled by STOV evolution, as we discuss later.

An important goal of this work is optimizing the laser pulsewidth to extend the filament length for applications [7,9]. Figure 2(d) plots the pulsewidth dependence of filament length $\Delta z_{fil}$ and length per unit pulse energy $\Delta z_{fil}/\varepsilon_{pulse}$. Here, $\Delta z_{fil}$ is defined as the energy deposition axial extent with $\partial_z \varepsilon_{dep} > 0.7$ µJ/cm, ~5 × the microphone noise floor; Fig. 2(a) illustrates $\Delta z_{fil}$ for a $\tau = 500$ fs pulse. It is seen that $\Delta z_{fil}$ increases with $\tau$, with a ~5.5 × increase from $\tau = 45$ fs to 2 ps and agreement between experiment and simulations. As confirmed by the simulations, the length increase is attributable to two effects: (1) enhanced self-focusing as the high-power time slices of longer pulses overlap with a stronger rotational response and (2) stronger refocusing cycles supported by longer pulses; this is discussed below in the context of STOV dynamics. Note from Fig. 2(d), however, that the ratio of filament length to initial pulse energy, $\Delta z_{fil}/\varepsilon_{pulse}$, increases only to $\tau \sim 200$ fs, while decreasing for larger $\tau$ as $n_{2,eff}$ remains constant. This is because the self-focusing strength increases with minimal additional pulse energy up to $\tau \sim 200$ fs, after which the near-saturation of $n_{2,eff}$ and the increasing contribution of avalanche ionization arrests self-focusing at lower intensities, limiting the energy deposition. Thus, while $\Delta z_{fil}$ increases with pulsewidth, efficiency decreases for $\tau > \sim 200$ fs.

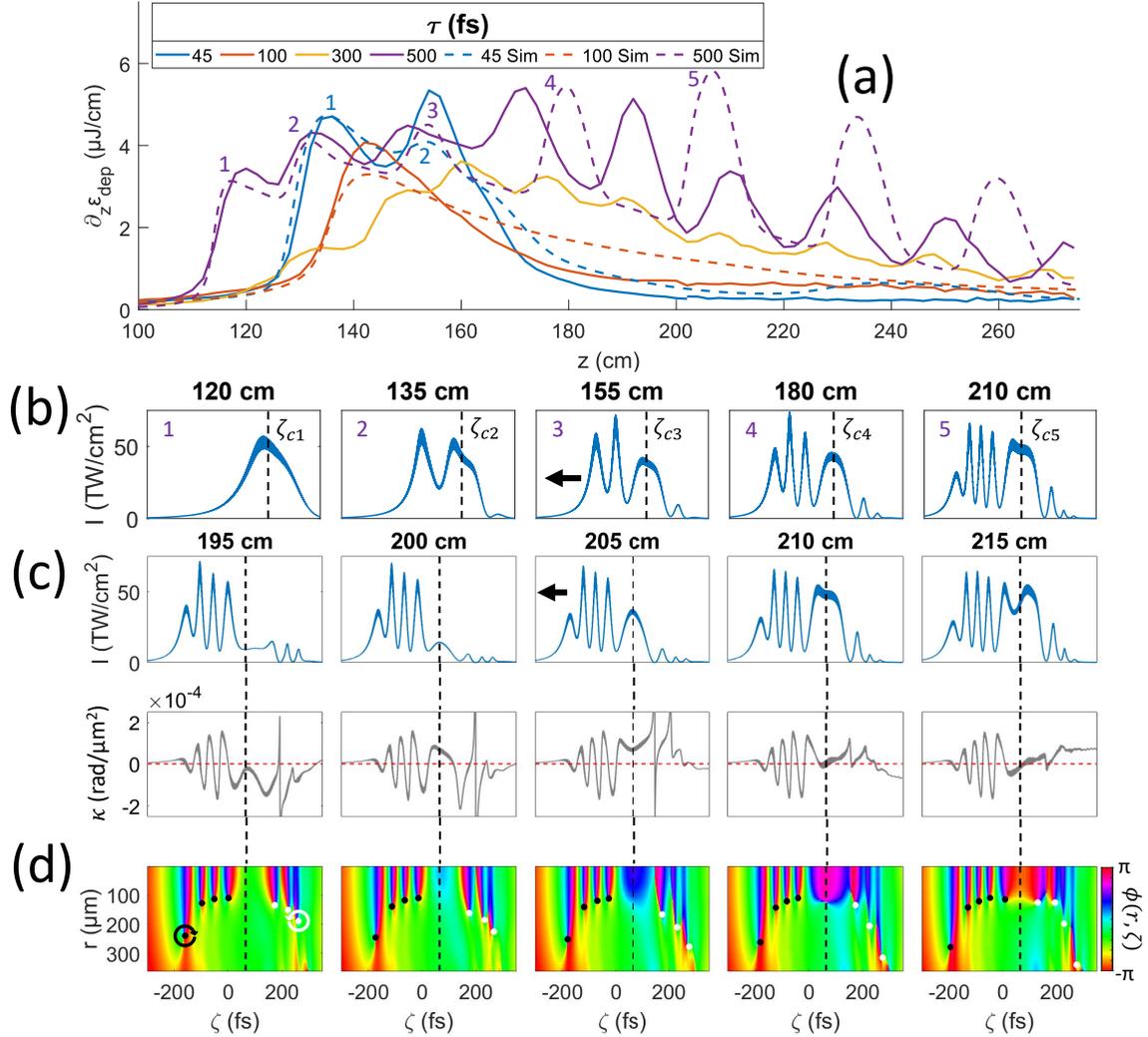

**Figure 3. (a)** Filament energy deposition profiles. Traces (solid curves) are stitched 100-shot averages of single shot traces aligned to the average collapse location. YAPPE simulations of energy deposition are plotted as dashed curves. **(b)** Simulated on-axis intensity envelope $I(r=0,\zeta;z)$ for the $\tau = 500$ fs pulse just before collapse arrest at the five $z$ locations corresponding to peaks 1-5 in panel (a). The black arrow marks the pulse propagation direction for all panels. **(c)** *Top row:* Simulated $I(r=0,\zeta;z)$ for the $\tau = 500$ fs pulse leading up to the 5$^{th}$ collapse arrest at $z \sim 210$ cm and local time $\zeta = \zeta_{c5}$, marked by the vertical dashed line. *Bottom row*: Simulated on-axis phase curvature $\kappa = \partial_r^2 \arg(E(r,\zeta;z))|_{r=0}$ calculated from simulation. **(d)** Spatiotemporal phase $\Phi(r,\zeta)$ corresponding to panels in (c), with $l = +1$ STOV phase singularities marked with black dots and those for $l = -1$ marked with white dots. The black and white arrows in the left panel indicate the direction of phase circulation.

We now examine the axially modulated energy deposition in more detail, elucidating the central role played by STOVs. Figure 3(a) plots axial energy deposition profiles (100 shot averages) for 45, 100, 300, and 500-fs pulses. The overlaid simulations reproduce the main qualitative features of the measurements: the longer pulses ($\geq 300$ fs) generate multiple deposition peaks whose amplitude increases with pulse duration. For the 100-fs pulse, the peaks are present but negligible compared to the baseline profile, while the 45-fs pulse has only 2 peaks near initial filament formation. In all cases, each deposition peak is caused by a single collapse arrest event. This is shown in Fig. 3(b) for the 500-fs pulse: near $z = 120$ cm, the on-axis pulse intensity envelope $I(r = 0, \zeta; z)$ reaches its peak just before collapse arrest at $\zeta = \zeta_{c1}$, generating deposition peak 1 (marked in Fig. 3(a)) and begins to split. The split pulse propagates to $z = 135$ cm, where collapse arrest at $\zeta = \zeta_{c2}$ leads to deposition peak 2 and another split. The pattern is that after the $N^{th}$ collapse arrest (and deposition peak $N$), the pulse has $N$ intensity peaks ahead of $\zeta = \zeta_{cN}$ and $N$ (weaker) peaks trailing it. As intensity peaks are added, they narrow and become increasingly packed into the front and back of the original pulse envelope.

To see how, say, the 5$^{th}$ deposition peak and related pulse structure are generated, Fig. 3(c) plots simulated $I(r = 0, \zeta; z)$ (upper row) and on-axis phase curvature $\kappa = \partial_r^2 \arg(E(r, \zeta; z))|_{r=0}$ (lower row), where $E$ is the complex electric field. The phase curvature includes the bound electronic and rotational contributions (focusing) plus the plasma and diffraction contributions (defocusing). As propagation proceeds, the pulse intensity at $\zeta = \zeta_{c5}$ (marked by the vertical dashed line) grows as $\kappa$ increases there due to the strong molecular lens, until collapse arrest begins by $z = 210$ cm as $\kappa$ crosses through zero and pulse splitting commences. At $z > 215$ cm, the pulse has developed 5 intensity peaks for $\zeta < \zeta_{c5}$ and 5 (weaker) peaks for $\zeta > \zeta_{c5}$. The role of STOVs is made clear by the corresponding spatiotemporal phase ($\phi(r, \zeta; z)$) plots in Fig. 3(d). Before the 5$^{th}$ deposition peak, 4 collapse arrest events have already taken place, each of which have launched a STOV pair with topological charges $l = \pm 1$, with the +1 STOV propagating forward in the moving window and the $-1$ STOV propagating backward (STOV pair generation and motion are discussed in detail in [14,15]). By the time the pulse has reached $z = 195$ cm, four +1 STOVs have accumulated at $\zeta < \zeta_{c5}$ (marked by black dots) and four $-1$ STOVs (marked by white dots) have accumulated at $\zeta > \zeta_{c5}$, where the rightmost $-1$ STOV has left the computation window owing to pulse steepening (see below). The STOV phase singularities (or topological defects) line up in $\zeta$ with the intensity peaks. For the $l = +1$ STOVs, $\kappa < 0$ before each singularity and $\kappa > 0$ after, and vice versa for the $l = -1$ STOVs. Each of the STOVs is a toroidal vortex whose singularity is wrapped around the pulse propagation (major) axis, with spacetime phase circulation around the singularity/null (minor) axis. Ref. [26] presents a video showing the full propagation evolution of $I(r = 0, \zeta; z)$, $\kappa$, and $\phi(r, \zeta; z)$ for the 500 fs pulse.

In Fig. 3, it is notable that the leading and trailing pulse arrays appear stable, with collapse arrest events occurring *only* at a single time slice in the gap between the arrays. This is because each of the STOV arrays already represents a self-consistent and regularized state of collapse-arrested nonlinear propagation. A new collapse can only take place at a time slice in the inter-array gap selected by the peak focusing strength of the molecular lens, which occurs several hundred femtoseconds after the pulse leading edge. As additional $\pm 1$ STOVs are generated, they are increasingly squeezed into opposite sides of the gap, forcing narrowing of the associated intensity peaks. At the front, like-sign vortex-vortex repulsion maintains peak separation, with the lead

STOV constrained from moving forward beyond the pulse envelope [14,15]. Similar dynamics takes place at the back of the pulse, except that the STOVs and intensity peaks are even more strongly compressed by self-steepening (owing to the Kerr-induced lower group velocity of earlier time slices), leading to the increased transverse separation of −1 STOVs and migration from the computation window owing to mutual repulsion (Fig. 3(d)).

The increasing amplitude of the deposition peaks and filament length with pulsewidth (at fixed $P/P_{cr}$) in Fig. 3(a) originates from the higher on-axis fluence and increased rotational response via $\Delta n_{b,rot}(t) = \int_{-\infty}^{t} dt' R(t-t') I(t')$, leading to stronger molecular focusing (as also seen in Fig. 2(c)). By contrast, the 45-fs pulse, which dominantly samples the instantaneous electronic nonlinearity, produces a shorter-range deposition profile with only 2 deposition peaks. For more detail, see Appendix A and the video in [26].

Measurements and simulations of STOV-mediated pulse envelopes for $z > 185$ cm are plotted in Fig. 4 for initial pulsewidths 50 fs − 400 fs. Here, the FROG setup of Fig. 1(d) was used to measure the mid-flight evolution of the filament optical core at a location well past the initial pulse collapse, so that multiple collapse arrest events have already occurred. The envelope for each initial pulsewidth was extracted from SHG-FROG images that were a composite of 10 shots. Considering the ~5 fs resolution of the SHG-FROG results, the experimental and YAPPE simulation plots are in qualitative agreement: for the 45-fs pulse, only two peaks develop, consistent with the dominant instantaneous nonlinearity at that pulsewidth and the deposition profile of Fig. 3(a). As the rotational contribution to the nonlinearity increases with pulsewidth, the number of peaks increases. In addition, self-steepening at the back of the pulse is increasingly evident as pulsewidth increases.

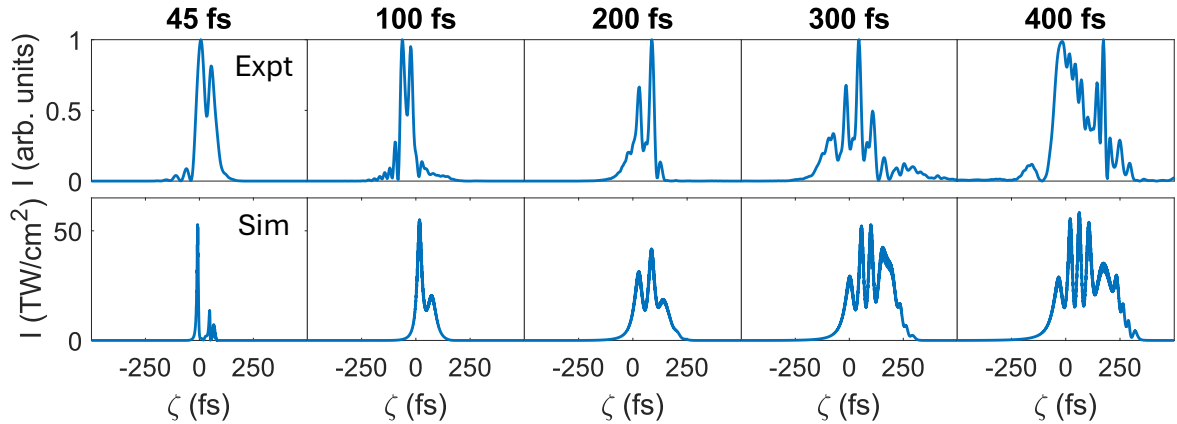

**Figure 4.** *Top row*: SHG-FROG measurements of the filament core at $z > 185$ cm for initial laser pulsewidths 45 fs-400 fs. An increasing number of peaks is generated for longer pulsewidths. In all cases, $P/P_{cr} = 5$. *Bottom row*: On-axis intensity envelope from YAPPE propagation simulations using the experimental parameters (pulsewidth, energy) to maintain $P/P_{cr} = 5$.

To summarize, we have demonstrated that long-pulse filamentation in air is best understood as a topological defect-driven nonlinear process whose macroscopic behavior is dictated by the repeated generation and evolution of spatiotemporal optical vortices. For pulses sufficiently long to sample the delayed rotational nonlinearity of air molecules, collapse arrest events occur periodically along the propagation path, each event generating a ±1 STOV pair and a peak in the axial energy deposition. The +1 and −1 charges migrate in opposite directions and accumulate into ordered arrays at the front and back of the pulse. These arrays mediate intrapulse energy flow and impose a stable regular multi-peak temporal envelope structure. The gap between the arrays acts as the source of new $\pm 1$ STOV pairs, with each collapse arrest cycle adding intensity peaks that become increasingly compressed within the evolving envelope. By contrast, short pulses dominated by the instantaneous electronic nonlinearity exhibit limited refocusing and significantly shorter propagation lengths. The delayed rotational response enhances effective self-focusing and stabilizes the dynamics, allowing longer pulses to produce longer filaments from more refocusing cycles and enabling long-range filament applications in the atmosphere.

The authors thank S. Hancock, A. Tartaro, and I. Larkin for technical assistance and discussions. This work was supported by the Air Force Office of Scientific Research (FA9550-25-1-0332) and the Department of Energy (DE-SC0024398).

*Appendix A: Propagation of 45 fs pulses in air*—Here we examine, via YAPPE simulation, the behavior of a pulse ($\tau = 45$ fs, $P/P_{cr} = 5.5$) much shorter than the air molecular rotational response, corresponding to the measured and simulated energy deposition profiles in Fig. 3(a) and pulse envelopes in Fig. 4 (for $P/P_{cr} = 5$) for $\tau = 45$ fs. The plots of $I(r = 0, \zeta; z)$ in the top row of Fig. 5(a) show two events (labelled 1 and 2) corresponding to the blue-labeled deposition peaks in Fig. 3(a). The first peak corresponds to the initial collapse arrest just past $z \sim 135$ cm, where $\pm 1$ STOVs are generated, with pulse splitting seen by 140 cm. The second peak is a remarkable consequence of the annihilation of the $-1$ STOV at $z \sim 155$ cm.

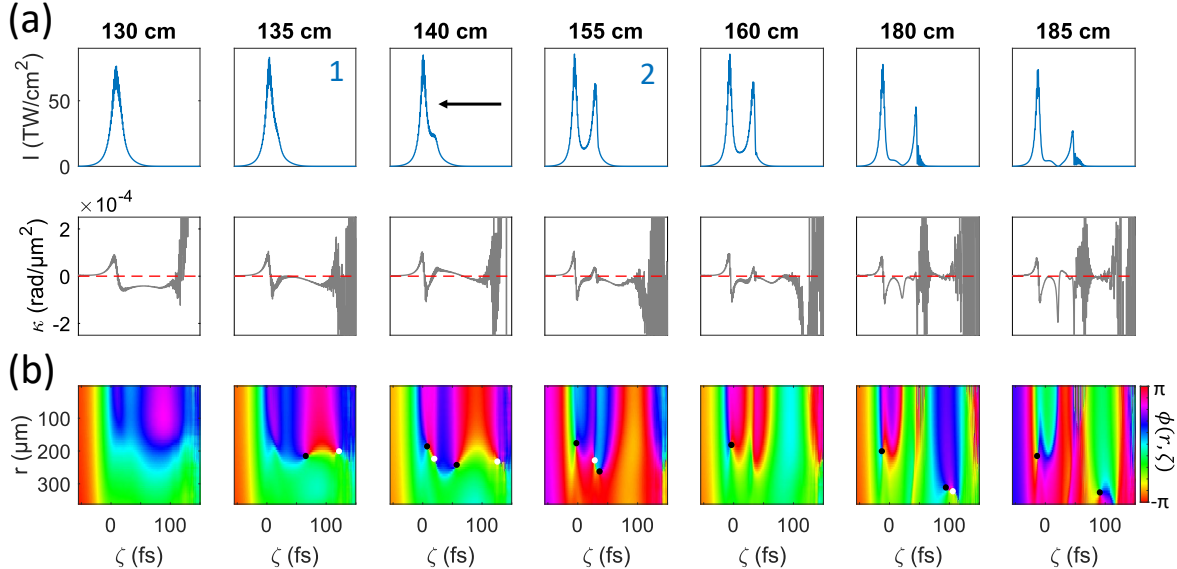

**Figure 5. (a)** *Top row*: Simulated on-axis intensity envelope $I(r = 0, \zeta; z)$ for $\tau = 45$ fs through filament propagation, with collapse arrest events labelled 1 and 2 matching the $z$ locations in 5(a). The black arrow marks the pulse propagation direction for all panels. *Bottom row*: On-axis phase curvature $\kappa = \partial_r^2 \arg(E(r, \zeta; z))|_{r=0}$ calculated from simulation. **(b)** Spatiotemporal phase $\phi(r, \zeta)$ corresponding to panels in (a), with $l = +1$ STOV phase singularities marked with black dots those for $l = -1$ marked with white dots.

As seen in Fig. 5(b), $\pm 1$ STOVs are generated slightly before $z \sim 135$ cm due to spatiotemporal phase shear from the rotational response peaking *behind* the pulse (a phenomenon discussed in [15]). Upon the first collapse arrest event, the associated $+1$ STOV follows the intensity peak and the $-1$ STOV moves backward. This $-1$ STOV is then annihilated near $z \sim 155$ cm by the forward moving rotationally-generated $+1$ STOV. Because electromagnetic energy *outflow* with respect to the propagation axis is mandated by the region between $\pm 1$ collapse arrest STOVs [14], the annihilation of the $-1$ STOV terminates this process, abruptly increasing the near-axis power and energy deposition. This gives rise to the second deposition peak. For longer pulses, since the rotational response overlaps substantially with the pulse itself, this STOV annihilation does not occur. (Rotational STOV pair generation is also seen at 180 cm, although at such a transverse extent as to not substantially impact filament propagation.)

# Supplementary Material: Topologically constrained high intensity light propagation

A. Goffin[1,2], L. Railing[1,3], G. Babic[1,3], and H. M. Milchberg[1,3,4*]

[1]*Institute for Research in Electronics and Applied Physics, University of Maryland, College Park, MD 20742*
[2]*Los Alamos National Laboratory, Los Alamos, NM 87545*
[3]*Dept. of Physics, University of Maryland, College Park, MD 20742*
[4]*Dept. of Electrical and Computer Engineering, University of Maryland, College Park, MD 20742*
* *milch@umd.edu*

## 1. Effective nonlinear index $n_{2,eff}$

In our experiments, we examine the pulsewidth-dependent laser field evolution and propagation track while keeping the ratio $P/P_{cr}$ constant; this ensures that all time slices of the pulse envelope initially experience a similar effective nonlinear index $n_{2,eff}$. We must therefore find the correct $n_{2,eff}$ that embodies the electronic plus rotational responses over the full pulse evolution. To do so, we consider the pulse's nonlinear phase pickup $\Delta\phi_{NL,b}(t)$ from bound electrons, whose intensity-weighted mean value over a pulse of FWHM duration $\tau$ is $\langle\Delta\phi_{NL,b}\rangle_\tau = \left(\int_{-\infty}^{\infty} I(t)dt\right)^{-1} k_0 L \int_{-\infty}^{\infty} I(t)\Delta n_b(t)dt$. Here $\Delta n_b(t) = n_2 I(t) + \int_{-\infty}^{t} dt' R(t-t')I(t')$ is the shift in bound electron refractive index from the instantaneous electronic response (first term) and delayed rotational response (second term), $R(t-t')$ is the diatomic molecular rotational response function [1–3], and $k_0$ and $L$ are the pulse central wavenumber and propagation distance. We then define a mean phase shift for an effective instantaneous nonlinearity with coefficient $n_{2,eff}$, $\langle\Delta\phi_{eff}\rangle_\tau = \left(\int_{-\infty}^{\infty} I(t)dt\right)^{-1} k_0 L \int_{-\infty}^{\infty} I(t)[n_{2,eff}I(t)]dt$. Setting $\langle\Delta\phi_{NL,b}\rangle_\tau = \langle\Delta\phi_{eff}\rangle_\tau$ yields

$$n_{2,eff}(\tau) = \left(\int_{-\infty}^{\infty} I^2(t)dt\right)^{-1} \int_{-\infty}^{\infty} I(t)\Delta n_b(t)dt, \tag{S1}$$

the effective nonlinear index for a pulse of width $\tau$.

## 2. Microphone array calibration

We first measured the total filament energy deposition via interferometric measurements of the filament-induced density depression, a technique developed in [4]. By directing the filament from air into the nozzle of a slow outflow helium cell [5,6] (see Fig. S1(a)), the filament is terminated mid-flight at the air-helium interface due to the low $n_2$ of helium (<$0.05 n_{2,air}$) [1,7]. The cell is mounted on a rail, enabling axially scannable mid-flight measurements of the filament. A weak co-propagating probe pulse ($\lambda_0 = 532$ nm, $\tau \sim 7$ ns) was directed along the filament path after a delay of 1 ms, picking up the axially integrated phase shift $\Delta\phi(x,y;z)$ from the residual thermal density depression ("density hole") left from filament heating of the air, where $(x,y)$ is a tranverse coordinate in the probe beam and $z$ is the location of the air-helium interface. The probe is imaged from the interface through a folded wavefront interferometer onto a CCD camera, as shown in Fig. S1(a), enabling extraction of $\Delta\phi(x,y;z)$ (Fig. S1(b)) by standard methods [8]. The integrated energy deposition up to location $z$ from the beginning of the filament is related to the phase shift by [4],

$$\varepsilon_{dep}(z) = -c_v T_0 \rho_0 k^{-1}(n-1)^{-1} \int \Delta\phi(x,y;z)dxdy \qquad (S2)$$

where $c_v = 717$ J/(kg K) is the isochoric specific heat of air, $T_0 = 293$ K is the ambient temperature of air, $\rho_0 = 1.225$ kg/m³ is the ambient density of air [9], $k = 2\pi/\lambda_0$ is the wave number of the probe and $n \approx 1 + 2.77 \times 10^{-4}$ is the ambient index of refraction of air for $\lambda_0 = 532$ nm.

To calibrate the microphone array, the integrated microphone signal, $\varepsilon_{mic}(z) = \alpha \sum_0^z S(z_i)$, up to position $z$ was fit to $\varepsilon_{dep}(z)$ for a range of $z$. Here, $S(z_i)$ is the signal of the microphone at $z = z_i$, and $\alpha$ is a coefficient determined by the best fit. Figure S1(c) plots $\varepsilon_{dep}(z)$ vs. $z$ along with the best fit $\varepsilon_{mic}(z)$ curve, which gives $\alpha = 14.3$ µJ/(cm V), providing the microphone array calibration.

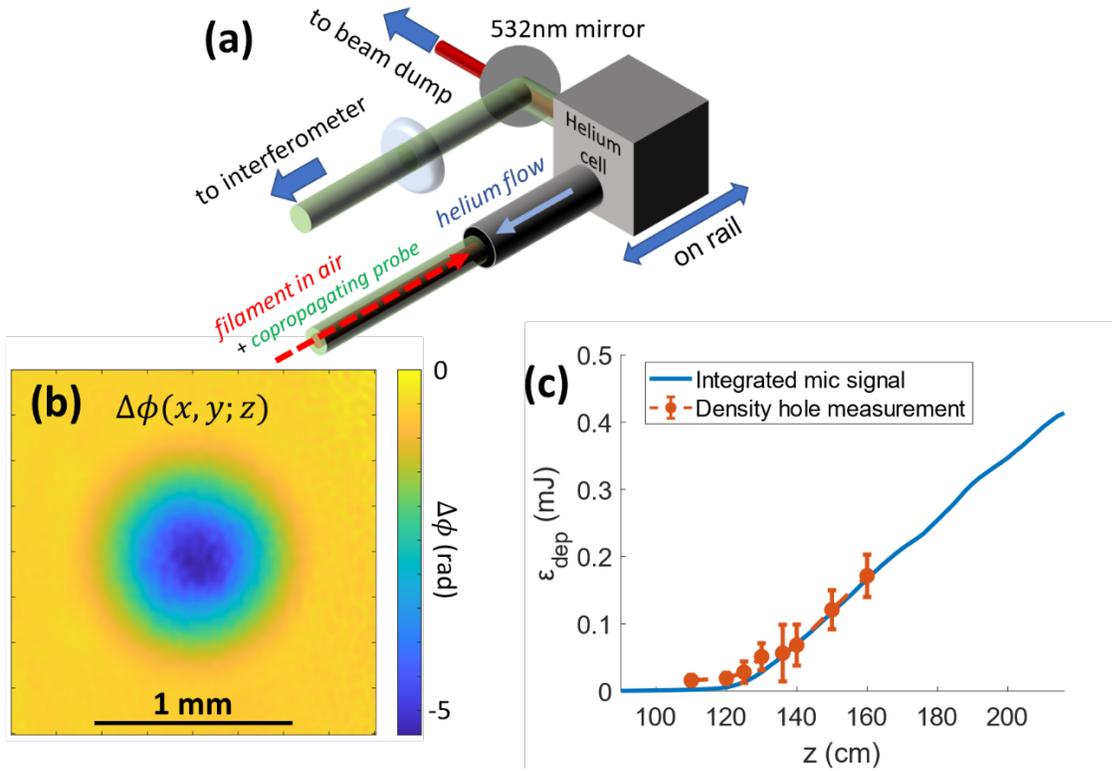

**Figure S1.** (a) Diagram showing probe ($\lambda_0 = 532$ nm, $\tau \sim 7$ ns) co-propagation with filament into the helium cell for interferometrically-based deposition measurement. (b) Total accumulated phase from a density hole generated by a $\varepsilon_{pulse} = 7.2$ mJ, $\tau = 500$ fs filament ($P/P_{cr} = 5.5$) at $z = 160$ cm, with 1 ms probe delay. (c) Integrated microphone signal $\varepsilon_{mic}(z)$ (blue curve) overlaid on interferometrically derived integrated energy deposition $\varepsilon_{dep}(z)$ (brown points). The best fit overlay gives a calibration of 1V (microphone signal) = 14.3 µJ/cm.

## 3. Propagation simulations using YAPPE

The simulations of beam propagation in this document were performed using a 2D+1 cylindrically-symmetric unidirectional pulse propagation equation (UPPE) [10] implementation called YAPPE ('Yet Another Pulse Propagation Effort'). This is a system of ordinary differential equations (ODEs) of the form

$$\frac{\partial}{\partial z} A_{k_\perp}(\omega, z) = i Q_{k_\perp}(\omega) 2\pi P_{k_\perp}(\omega, z) e^{-i\left(k_z - \frac{\omega}{v_g(\omega_0)}\right)z} . \quad (S3)$$

In Eq. (S3), $A = A_{k_\perp}(\omega, z)$ is an auxiliary field related to the Fourier transform of the optical field by $E = A e^{ik_z z}$, $\omega$ is angular frequency, $v_g(\omega_0)$ is the group velocity of the pulse at central frequency $\omega_0$, $k_z = ((\omega/v_g(\omega))^2 - k_\perp^2)^{1/2}$ is the longitudinal (z-direction) wavenumber, $Q_{k_\perp}(\omega) = \omega/ck_z$, and $P_{k_\perp}(\omega, z)$ is the nonlinear polarization of the medium including third-order electronic nonlinearities in air [1], the molecular rotational response [1–3], the plasma response, and loss mechanisms [11]. The spectrum of radial spatial frequencies $k_\perp$ indexes a system of ordinary differential equations, which is solved using a GPU implementation of MATLAB's ODE45 function. The ionization rate is computed using a "strong field" module best suited for our Keldysh parameter $\gamma \sim 1$ and includes a Drude avalanche ionization model for > 1 ps FWHM pulses. The electron density therefore evolves as

$$\frac{\partial N_e(r, \zeta)}{\partial \zeta} = \frac{\sigma}{U_I} N_e I(r, \zeta) + N_g \Gamma_{PI}\{I(r, \zeta)\}, \quad (S4)$$

where $\zeta = t - z/v_g$ is local time in the moving window, $\sigma$ is the Drude collision cross-section, $U_I$ is the effective ionization energy of air, $N_g = 2.5 \times 10^{19} cm^{-3}$ is the number density of air, and $\Gamma_{PI}\{I(\zeta)\}$ is the photoionization rate [12] as a function of time. Eq. (S4) has an analytic solution that is used in YAPPE to calculate $N_e(r, \zeta)$:

$$N_e(r, \zeta) = N_g e^{\frac{\sigma}{U_I} \int_0^\zeta I(r, \zeta') d\xi'} \int_0^\zeta \Gamma_{PI}\{I(r, \zeta')\} e^{-\frac{\sigma}{U_I} \int_0^{\zeta'} I(r, \zeta'') d\zeta''} d\zeta' . \quad (S5)$$

Energy deposition is determined from the calculated fields by three mechanisms: photoionization, molecular rotation, and electron-neutral collisions (inverse Bremsstrahlung), and is computed as

$$\partial_z \varepsilon_{dep} = 2\pi \iint \left( \left[ \frac{\partial N_e^{N_2}}{\partial \zeta} U_I^{N_2} + \frac{\partial N_e^{O_2}}{\partial \zeta} U_I^{O_2} \right] + \frac{1}{c} \frac{\partial \Delta n_{rot}}{\partial \zeta} I + \frac{\partial N_e}{\partial \zeta} U_p + \nu_{en} N_e U_p \right) r dr d\zeta, \quad (S6)$$

where $N_e^X(r, \zeta)$ is the electron density in the filament contributed by air species $X$, $U_I^X$ is the ionization energy of air species $X$, $\Delta n_{rot}(r, \zeta) = \int_{-\infty}^\zeta d\zeta' R(r, \zeta - \zeta') I(r, \zeta')$ is the time-dependent index shift from molecular rotations, where $R(\zeta)$ is the rotational response function [1–3], $I(r, \zeta) = \frac{c}{8\pi} E^2(r, \zeta)$ is the laser field intensity, $\nu_{en} = 2.86$ ps$^{-1}$ is the electron-neutral collision rate [13], and $U_p = e^2 E^2 / 4m\omega_0^2$ is the ponderomotive energy of a free electron in the plasma. In Eq. (S6), the first term (bracketed) through the fourth term correspond, respectively, to laser energy deposition into field ionization, molecular rotations, ponderomotive kinetic energy imparted to newly freed electrons, and electron-neutral collisional heating.

## 4. Ionization and periodic refocusing

While our experiments and simulations are in good qualitative agreement, the measured distance between energy deposition peaks (or refocusing cycles) is 20-25 cm, while simulations give a period of ~27 cm (see Fig. 2(a)). This discrepancy can be explained by sensitivity to model details, which we illustrate by comparing YAPPE simulation results using two ionization models, the ADK ionization model [14] and the strong field ionization model [12] (used throughout the rest this paper). For a $\tau = 400$ fs pulse (and $P/P_{cr} = 5.5$), Fig. S2(a) plots the calibrated microphone trace, overlaid by the simulated energy deposition profile using the strong field ionization model. Here, the simulated deposition peaks and their separation are mostly higher than in the experiment. Figure S2(b) plots the same experimental trace, now overlaid by a simulation using the ADK ionization model. Here, the simulated deposition peaks are now lower than in the experiment, and their spacings are comparable. While there may be other sources of sensitivity in our modeling, such as assuming a rigid diatomic rotor for the rotational response function $R(\zeta)$ of air constituents, and the particular choice of avalanche model at longer pulsewidths, it is likely that the most sensitive choice is that of photoionization model, owing to the exponential dependence of ionization rates on laser fields and ionization potentials, as well as estimated sensitivity to electron density and transverse size found in Sec. 5 below. Given that our Keldysh parameter is $\gamma \sim 1$, we chose the strong field ionization model [12] for our propagation simulations.

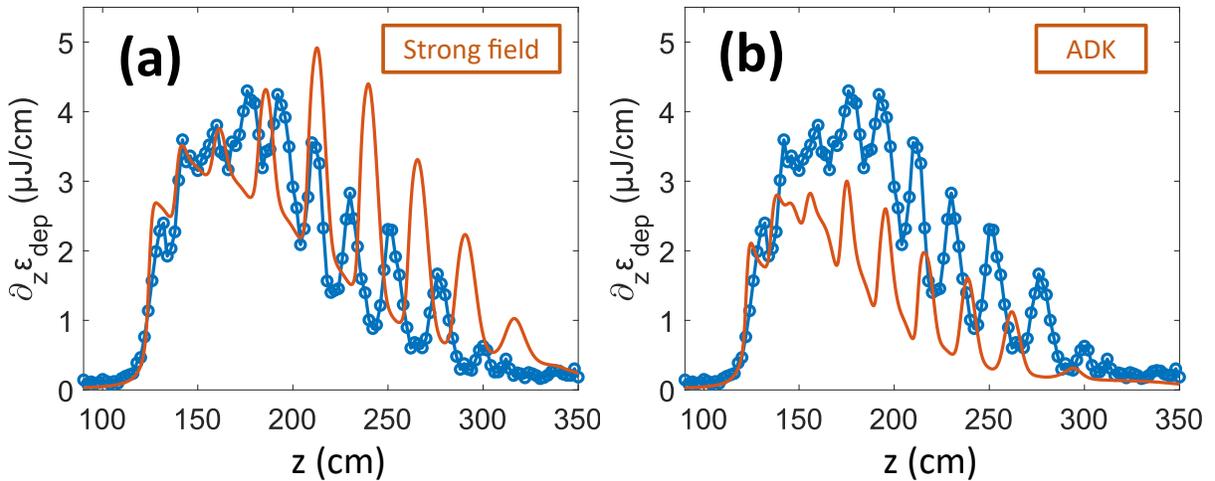

**Figure S2. (a)** Calibrated microphone trace (blue circles and curve) and YAPPE simulation of filament energy deposition (brown curve) using the strong field ionization model [12]. **(b)** Calibrated microphone trace (blue circles and curve) and YAPPE simulation of filament energy deposition (brown curve) using the ADK ionization model [14].

## 5. Periodic collapse arrest and energy deposition

Here we present a simple model for the periodic collapse arrest (refocusing cycles) observed in atmospheric filamentation of "rotationally long" ($\tau > \sim 150$ fs) pulses. We assume that pre-collapse arrest of an envelope time slice in the gap between the STOV arrays, the intensity profile is $I(r,\zeta) = I_0(\zeta)e^{-2r^2/w^2}$ and that the field in the gap experiences an electron density profile $N_e(r,\zeta) = N_{e0}e^{-2r^2/w_e^2}$ generated by the leading pulse array. The refractive index is $n(r,\zeta) = n_0 + \Delta n(r,\zeta)$, with the perturbation $\Delta n(r,\zeta) = n_{2,eff}I(r,\zeta) - N_e(r,\zeta)/2N_{cr}$. Near the optical axis for $r \ll w, w_e$ we have

$$\Delta n(r,\zeta) \approx \Delta n(0,\zeta) - 2r^2\left(\frac{n_{2,eff}I_0(\zeta)}{w^2} - \frac{N_{e0}}{2w_e^2 N_{cr}}\right) = \Delta n(0,\zeta) - ar^2, \quad (S7)$$

modeling the system as an effective graded index waveguide, where $\Delta n(0,\zeta) = \Delta n(0) = n_{2,eff}I_0 - N_{e0}/2N_{cr}$. As the plasma diameter is many optical wavelengths, we can approximate propagation using the ray evolution equation [15], where $r$ is transverse position of an optical ray,

$$\frac{d}{dz}\left(n\frac{dr}{dz}\right) = \frac{dn}{dr}. \quad (S8a)$$

This yields, for the refractive index $n = n_0 + \Delta n(0) - ar^2$,

$$\frac{d^2r}{dz^2} + \left(\frac{2a}{n_0 + \Delta n(0)}\right)r = 0. \quad (S8b)$$

The refocusing period is then

$$\Delta z_{ref} = \pi\sqrt{(n_0 + \Delta n(0))/2a} = \frac{\pi}{\sqrt{2}}\sqrt{\frac{n_0 + \Delta n(0)}{2n_{2,eff}I_0/w^2 - N_{e0}/w_e^2 N_{cr}}}, \quad (S9)$$

where we consider $I_0$ and $N_{e0}$ in the gap region. Note that Eq. (S9) can also be derived as $\Delta z_{ref} = \pi/|\beta_{p=1,m=0} - \beta_{p=0,m=0}|$, where $\beta_{p,m} = k_0 n_0 - \sqrt{2a/n_0}(2p + |m| + 1)$ is the wavenumber of the $(p,m)$ Laguerre-Gaussian TEM mode solution for the effective quadratic index of Eq. (S7), where $p = 0,1,2,...$ and $m = 0, \pm 1, \pm 2,...$ are the radial and azimuthal indices.

We substitute into Eq. (S9) some results from the YAPPE simulations (noting that $n_0 + \Delta n(0) \approx 1$). From Fig. 3(c) of the main paper, the average intensity in the gap leading to subsequent collapse is $\sim 4 \times 10^{13}$ W/cm$^2$. Using $n_{2,eff} \sim 4 \times 10^{-19}$ cm$^2$/W from Fig. 1(a), and $w = 100$ $\mu m$ and $w_e = 50$ $\mu m$ from the simulations, we get $\Delta z_{ref} \sim 20$ cm (comparable to the period in Figs. 2 and 3) if we choose $N_{e0} > \sim 10^{16}$ cm$^{-3}$. This density is in line with the simulations and previous measurements of filament electron density generated by few hundred fs pulses [16]. In Eq. (S9), note the sensitivity of $\Delta z_{ref}$ to the difference in terms in the right-side denominator, a sensitivity determined mainly by the ionization model leading to $N_{e0}$ and $w_e$. This is consistent with the discussion in Sec. 4 above.

## 6. Videos of long pulse and short pulse propagation

Propagation of $\tau = 500$ fs pulse

Propagation of $\tau = 45$ fs pulse